%% file: beamdamage.tex
\documentclass[journal=ancac3,manuscript=article,layout=twocolumn]{achemso}

\usepackage{chemformula} 
\usepackage[T1]{fontenc} 
\usepackage{siunitx}
\usepackage{amsmath}
\usepackage{adjustbox}
\usepackage{subcaption} 
\usepackage{stackengine} 

\title{Beam induced heating in electron microscopy modeled with machine learning interatomic potentials}

\author{Cuauhtemoc Nu$\tilde{\text{n}}$ez Valencia}
\affiliation[DTU Physics]{Computational Atomic-scale Materials Design (CAMD), Department of Physics,
  Technical University of Denmark, DK-2800 Kgs.\  Lyngby, Denmark}
\author{William Bang Lomholdt}
\affiliation[DTU Nanolab]{National Centre for Nano Fabrication and Characterization,
  Technical University of Denmark, DK-2800 Kgs.\  Lyngby, Denmark}
\author{Matthew Helmi Leth Larsen} 
\affiliation[DTU Physics]{Computational Atomic-scale Materials Design (CAMD), Department of Physics,
  Technical University of Denmark, DK-2800 Kgs.\  Lyngby, Denmark}
\author{Thomas W. Hansen}
\affiliation[DTU Nanolab]{National Centre for Nano Fabrication and Characterization,
  Technical University of Denmark, DK-2800 Kgs.\  Lyngby, Denmark}
\author{Jakob Schiøtz}
\affiliation[DTU Physics]{Computational Atomic-scale Materials Design (CAMD), Department of Physics,
  Technical University of Denmark, DK-2800 Kgs.\  Lyngby, Denmark}
\email{schiotz@fysik.dtu.dk}


\begin{document}


\begin{abstract}
    We develop a combined theoretical and experimental method for estimating the amount of heating that occurs in metallic nanoparticles that are being imaged in an electron microscope.  We model the thermal transport between the nanoparticle and the supporting material using molecular dynamics and eqivariant neural network potentials.  The potentials are trained to Density Functional Theory (DFT) calculations, and we show that an ensemble of potentials can be used as an estimate of the errors the neural network make in predicting energies and forces.  This can be used both to improve the networks during the training phase, and to validate the performance when simulating systems too big to be described by DFT. The energy deposited into the nanoparticle by the electron beam is estimated by measuring the mean free path of the electrons and the average energy loss, both are done with Electron Energy Loss Spectroscopy (EELS) within the microscope.  In combination, this allows us to predict the heating incurred by a nanoparticle as a function of its size, its shape, the support material, and the electron beam energy and intensity.
\end{abstract}



\section{Introduction}

In High-Resolution Transmission Electron Microscopy (HR-TEM) the sample is irradiated with high energy electrons that move through the sample, interacting with it to form the final image.  In order to have a good time resolution and sufficient signal-to-noise ratio (S/N) a high electron dose rate is 
needed, typically in the order of 
$10^3 - 10^5 \text{e}^-/\text{Å}^2 \text{s}$). However, this unavoidably results in interactions with the sample, mainly in the form of beam damage and localized heating \cite{VanDyck2015DoHREM}.

In this paper we show how molecular dynamics (MD) simulations using
machine learning potentials can be used to quantify the amount of
heating incurred by the sample due to effects of the beam.  As a model
system, we choose gold nanoparticles supported on hexagonal boron
nitride (hBN), where we also provide experimental data on the energy
deposition into the nanoparticle.  We illustrate the generalizability
of the method by also applying it to titanium dioxide (\ch{TiO2})
supports.

Metallic nanoparticles supported by metal oxides are widely used in heterogeneous catalysis. Electron microscopy is often used to study these kind of systems, but the effect of the electron beam on the sample is difficult to quantify. Model catalysts based on Au nanoparticles on Titanium dioxide (\ch{TiO2}) supports are actively studied, both to understand the relation between shape, size, and catalytic activity \cite{Bond2010SourceNanoparticles,Falsig2008TrendsNanoparticles,Brodersen2011UnderstandingSimulations}, and to study and quantify for the effects of the beam.  Experimentally, there have been attempts to measure the local heating for example through the shift in the plasmon energy due to thermal expansion \cite{Mecklenburg2015NanoscaleDevices}.

There have also been a number of studies of the local heating of nanoparticles in the electron beam\cite{Kryshtal2022EffectTEM,Jose-Yacaman2005SurfaceNanoparticles,Egerton2004RadiationSEM,Gryaznov1991RealBeams}, but it is difficult to model both the interactions between beam and nanoparticle, and the heat flow away from the nanoparticle.  

During the imaging process, the beam electrons interact elastically and inelastically. In the latter case, these interactions result in the deposition of energy in various forms such as plasmons, excitons, electron hole-pairs, etc. These excitations in the nanoparticle decay relatively quickly (in the range of femtoseconds to picoseconds) compared to the electron dose rate and contribute to the generation of heat.  This heat has to be transported away from the nanoparticle, mainly through thermal conductivity by the support.  
This heat transport is to  a large degree limited by heat flow through the interface between the nanoparticle and the support, where the mismatch of phonon frequencies limits the rate of heat flow.  

To quantify this, we perform molecular dynamics (MD) simulations of the heat flow.  
Due to the large size of these systems (several nanometers),
calculations based on Density Functional Theory (DFT) is limited by
computational capacity.  While realistic interatomic potentials are
available for most metals \cite{Tadmor2011TheModels}, the quality of
interatomic potentials for oxides are limited, and the matching
between the metallic nanoparticle and the oxide support can typically
not be described by interatomic potentials.  For this reason, we fit
Machine Learning interatomic potentials to the relevant systems, this
enables us to make medium to large scale molecular dynamics
simulations with almost the accuracy of DFT, but with a computational
cost that is orders of magnitude smaller.

We chose the recently proposed 
 E(3)-equivariant machine learning potentials of Batzner et
 al. \cite{Batzner2022E3-equivariantPotentials} as implemented in the
 NequIP package \cite{SimonBatznerNequIP}.  It implements a E(3)-equivariant neural network \cite{e3nn_paper} approach for learning interatomic potentials from ab-initio calculations for molecular dynamics simulations.  It has shown good performance and accuracy in the field of machine learning potentials, while needing fewer data to produce the same or better results as machine learning potentials based on invariant descriptors \cite{Batzner2022E3-equivariantPotentials,Zhao2023HighPredictions,Bunting2023ReactivityPropane}.

\section{Results and discussion}

 
 We employ the NequIP package to create two interatomic potentials: one tailored for Au on \ch{TiO2}, and another designed for Au on hBN. It's important to note that these potentials are not intended to be universally applicable to all systems involving these three elements. Specifically, the Au/\ch{TiO2} potential is not suitable for predicting behaviors in unrelated systems such as gold oxides or metallic titanium-gold alloys. 

For the Au/\ch{TiO2} potential, we used a data set of around 3800 DFT
calculations that contains diverse configurations of bulk Au, bulk
\ch{TiO2}, Au Nanoparticles (Au-NP), \ch{TiO2} surfaces, and
Au-nanoparticles on \ch{TiO2} surfaces.  Both the anatase and rutile
crystal structures were used for \ch{TiO2}.  Similarly, a data set was
made with around 1325 DFT calculations, containing Au bulk, bulk hBN,
hBN layers, Au nanoparticles on hBN. 


After hyperparameter optimization, the resulting network has the parameters given in table \ref{tab:parameters},  and produces RMS errors in the forces and energies of respectively \SI{0.071}{\eV\per \angstrom} and \SI{0.11}{\eV\per atom}.  The Supplementary Online Information (SOI) Figure \ref{fig:learningcurve} shows a learning curve, i.e. the quality of the prediction as a function of the training set size.

\begin{table}[tpb]
    \centering
    \begin{tabular}{|c|c|c|}
        \hline
        \textbf{Parameter}  & \textbf{ $TiO_2$} & \textbf{ $hBN$}\\
        \hline
        cutoff [\si{\angstrom}] & 5 &5 \\
        \hline
        No. of vectors & 32 & 32\\
        \hline
        No. of scalars & 32 & 32\\
        \hline
        Training size & 3188  &  1150\\
        \hline
        Validation size & 797  & 125\\
        \hline
        Batch size & 7  & 7\\
        \hline
        Epochs & 200  & 200\\
        \hline
        Weight on forces & 50  & 50\\
        \hline
        Weight on forces & 50  & 50\\
        \hline
        
        \textbf{F-RMSE} [\si{\eV\per \angstrom}] & 0.071 & 0.045\\
        \hline
        \textbf{E-RMSE} [\si{\eV\per atom}] & 0.011  & 0.014\\
        \hline
    \end{tabular}
    \caption{Parameters of the network, values for 
 the error in the forces and energies shown at the bottom}
    \label{tab:parameters}
\end{table}

In order to validate the accuracy of predictions made by a neural network on large systems that cannot be calculated using Density Functional Theory (DFT), we correlated the accuracy with the uncertainty of an ensemble comprised of $M$ networks trained with the same hyperparameters and on the same training set; we use $M = 5$ unless otherwise noted.  To evaluate the uncertainty, we take a trajectory file from a Molecular Dynamic (MD) simulation and calculate the forces at each time step for each atom using the M different networks.  The MD simulation is performed using the arithmetic ensemble average of the forces and energies from the $M$ networks, while keeping track of the variation within the ensemble.  If the model is performing well, we expect both low disagreement between the networks and DFT, as well as low uncertainty between the $M$ networks. To evaluate the accuracy of the model we stick to the Root Mean Square Error magnitude (RMSEmag) as recommended by Morrow \emph{et al.}\cite{Morrow2023HowPotentials}.

The error in the prediction of the force is defined as
\begin{equation}
    \delta F_k = \left| \vec F_k^{\text{pred}} - \vec F_k^{\text{DFT}}\right|
\end{equation}
where $F_k^{\text{pred}}$ is the ensemble-averaged predicted force on the $k$-th atom, and $F_k^{\text{DFT}}$ is the corresponding force from DFT.  Note that $\delta F$ is a scalar.  The variance of the force is calculated as
\begin{equation}
    \sigma F_k = \sqrt{ \frac{1}{M}\sum_{j=1}^M \left| \vec F_{j,k} - \vec F_k^{\text{pred}} \right|^2}
\end{equation}
where $\vec F_{j,k}$ is the force on atom $k$ as predicted by the $j$-th member of the ensemble of neural networks.  Thus $\sigma F_k$ measures the disagreement amongst the neural networks, while $\delta F_k$ measures their error compared to DFT.

\begin{figure}[tbp]
    \centering
    \includegraphics[width=0.5\textwidth]{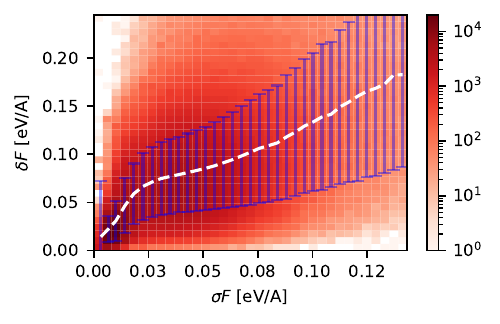}
    \caption{2D histogram plot of the error in the forces as a function of the uncertainty, when the uncertainty between the networks increases the error also increases.  Each atom in the data set produce a data point in this graph.  The x-axis is the variance between the neural networks in the ensemble ($\sigma F$), the y-axis is the error with respect to DFT ($\delta F$). The error bars show the standard deviation of each column, while the dotted line represents the average value of each column.}
    \label{fig:error-vs-variance}
\end{figure}
The results from the ensemble shown in Figure \ref{fig:error-vs-variance} demonstrate the relationship between the uncertainty of the network and the error in the predictions. As the uncertainty increases, the error in the predictions also increases, indicating that the network is struggling to understand the interactions between the atoms.  The variance between networks can thus also be used to gauge the error in simulations that are too large to validate with DFT simulations of snapshots.  The ability to correlate the uncertainty with the error furthermore allows for greater insight into the limitations of the network and areas where further improvement may be needed.

\begin{figure}[tbp]
    \centering
    \includegraphics[width=0.9\linewidth]{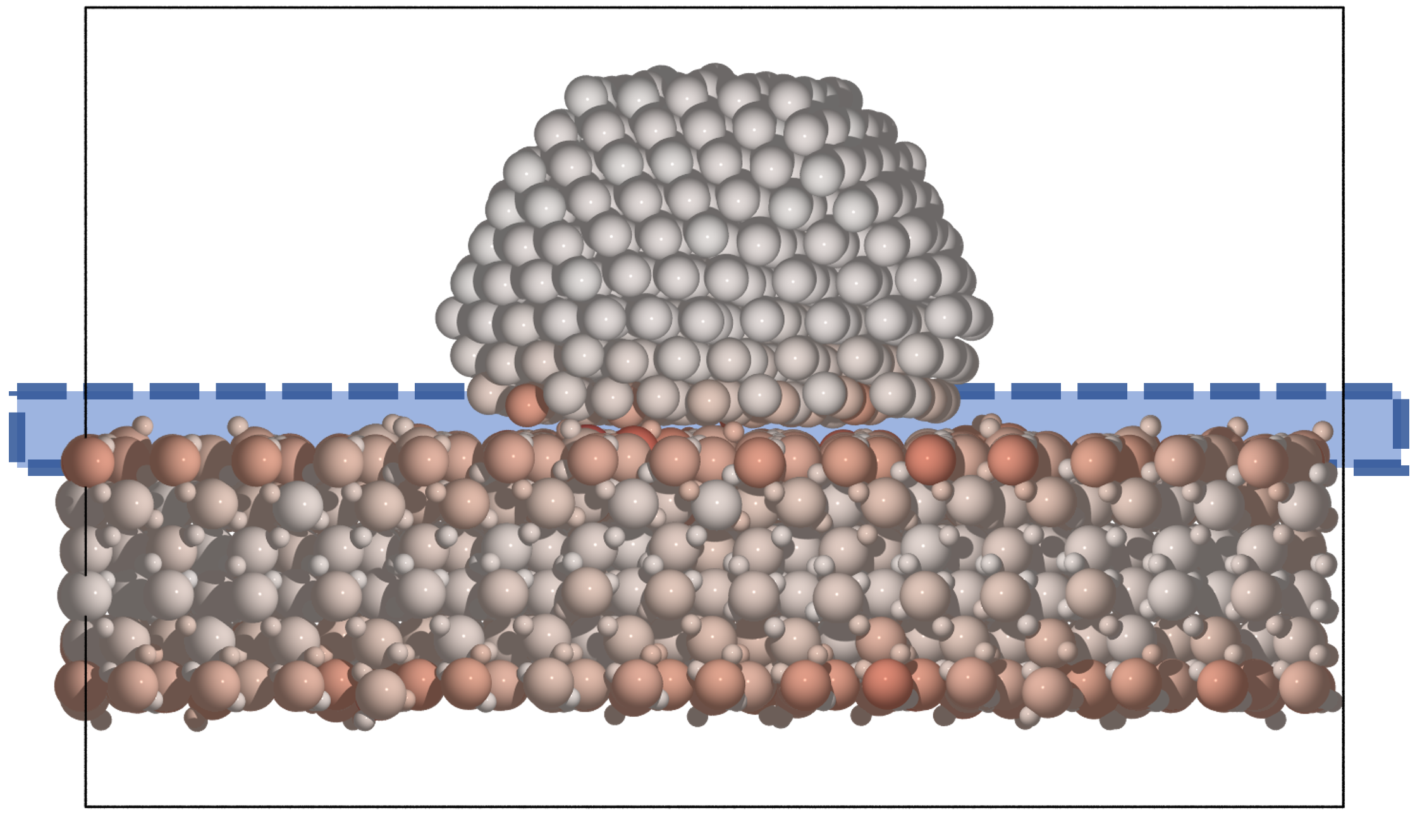}\\
    \includegraphics[width=\linewidth]{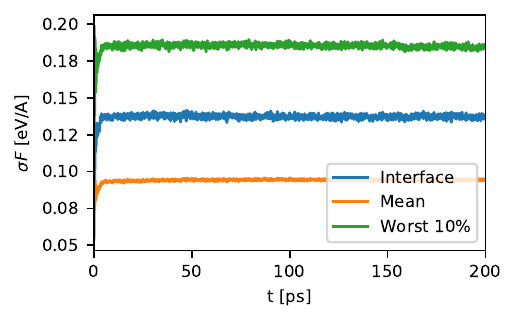}
    \caption{Where does the network struggle to predict the forces?  In the top panel, we visualize a gold nanoparticle positioned on a \ch{TiO2} substrate. Here, we've identified the top 10\% of atoms with the highest variance in forces and color-coded them based on this variance; a deeper shade of red indicate higher $\sigma F$ values.  The blue box indicates the interface region. Below the mean values of $\sigma F$ as a function of time for all the atoms (orange), the interface atoms (blue) and the 10\% worse atoms (green).  The majority of the 10\% worse atoms are in the interface.}
    \label{fig:interface}
\end{figure}

For example, we realized that many of the atoms with a large value of
$\sigma F$ are located at the interface between Au and \ch{TiO2}, this
can be seen in Figure \ref{fig:interface}.  The $\sigma F$ values of
the atoms within the interface are on average twice as large as the
average, we believe this is because of the complexity of the
interactions of oxygen atoms with both gold and titanium, causing the
network to struggle to predict forces accurately.

Once a machine learning potential has been fitted to the relevant
systems, molecular dynamics can be used to estimate the heat transfer,
as described in the Methods section.

The heat input rate into the nanoparticle from the electron beam can be approximated using the following equation \cite{Egerton2004RadiationSEM}:
\begin{equation}
    \label{eq:qin}
    \dot Q_{in} = S D \langle E \rangle \frac{t}{\lambda}
\end{equation}
where $S$ is the cross-sectional area of the nanoparticle in the beam,
$D$ is the dose rate, $\langle E \rangle$ is the average energy
deposited by an electron \emph{if} it is scattered, $t$ is the average
thickness of the nanoparticle in the direction of the beam, and
$\lambda$ is the mean free path of the electrons in the nanoparticle.
We obtain typical values for $\langle E \rangle$ and $\lambda$ from
Electron Energy-Loss Spectroscopy (EELS) as described in the Methods
section; for Au nanoparticles exposed to a
\SI{300}{\kilo\electronvolt} electron beam, we find
$\langle E \rangle \approx \SI{43.6}{\electronvolt}$ and
$\lambda \approx \SI{104}{\nano\meter}$.

The energy absorption density is much lower in the substrate as the lower atomic number ($Z$) results in a larger mean free path for the electrons ($\lambda \approx Z^{-1/2}$) \cite{Iakoubovskii2008MeanBehavior}. Furthermore, not all the substrate is illuminated by the beam, so we assume that the temperature of the substrate is the close to the operating temperature of the microscope.

In equilibrium, the heat flow into the nanoparticle must match the heat flow between the nanoparticle and the substrate
\begin{equation}
    \label{eq:qout}
    \dot Q_{out} = k A \Delta T
\end{equation}
where $k$ is a thermal conductance per unit area (its reciprocal is often denoted the R-value), $A$ is the area of the interface between the nanoparticle and the support, and $\Delta T$ is the temperature difference.  Note that the areas $S$ and $A$ are often not the same, their ratio depend on the shape of the nanoparticle (Supplementary Online Information (SOI) Figure \ref{fig:SAratio}).  The temperature rise of the nanoparticle thus becomes
\begin{equation}
    \Delta T = \frac{S D \langle E \rangle}{k A} \frac{t}{\lambda}
\end{equation}

Here we are assuming that the transfer of heat between the
nanoparticle and the substrate is significantly slower than heat
conductivity within both the nanoparticle and the substrate, that is
easily verified to be the case (see Methods).  This differs from at least some previous studies, which assumed that the heat transfer within the nanoparticle was rate limiting \cite{Gryaznov1991RealBeams}.

The thermal conductance $k$ in equation (\ref{eq:qout}) is estimated using molecular dynamics for the two cases of an Au nanoparticle on a \ch{TiO2} substrate and on a substrate of hexagonal boron nitride (hBN).  Nanoparticles are created with different sizes and different shapes, to vary both their volumes and the ratio between the beam cross-section ($S$) and the contact area to the support ($A$).  In all cases, the nanoparticle is oriented with (111) planes parallel to the support, as that is how they are usually orientated when observed in the electron microscope \cite{Yuan2021InOxidation,Liu2019TransformationsMicroscopy}.

The temperature difference between the nanoparticle and the substrate is tracked throughout the simulation, and fitted to the exponential decay expected from Eq. (\ref{eq:qout}):
\begin{equation}
    \Delta T(t) = \Delta T(0) \exp\left(- \frac{k A}{N c_p} t\right)
\end{equation}
where $N$ is the number of atoms in the nanoparticle, and $c_p$ its
atom-specific heat capacity.  It is found from the simulations to be
approximately \SI{2.9e-4}{\electronvolt\per\kelvin} in good agreement
with the experimental value of  
\SI{2.63e-4}{\electronvolt\per\kelvin} \cite{Jiang2013Size-DependentNanoparticles}, and with the
value expected in a harmonic potential, $c_p \approx c_v \approx 3 k_B
= \SI{2.58e-4}{\electronvolt\per\kelvin}$.  The deviation
is most likely due to anharmonic effects of the surface atoms, as we
find a value of \SI{2.55e-4}{\electronvolt\per\kelvin} for bulk gold.

From the simulations, we get average values of $k$ for the two
substrates:
$k_{\text{TiO}_2} =
\SI{5.1e-4}{\electronvolt\per\nm\squared\per\pico\second\per\K}$ and
$k_{\text{hBN}} =
\SI{3.7e-4}{\electronvolt\per\nm\squared\per\pico\second\per\K}$.  For
\ch{TiO2}, we use only anatase (101) surfaces values because they are
the most stable and lowest in energy \cite{martsinovich2012tio}, the
$k$-value for the rutile surfaces are approximately a factor 2 higher.
This confirms the assumption that the majority of the temperature
difference is over the interface.  With an experimental value for the
heat conductivity of gold of $\SI{317}{\watt\per\meter\per\kelvin} $,
a nanoparticle would need a thickness of 3.7 $\mu$m for the
temperature drop across the nanoparticle to be similar to the
temperature drop across the interface, confirming that we are in the
regime where the interface dominates the heat transport.

\begin{figure}[tbp]
  \centering
  \includegraphics[width=0.9\linewidth]{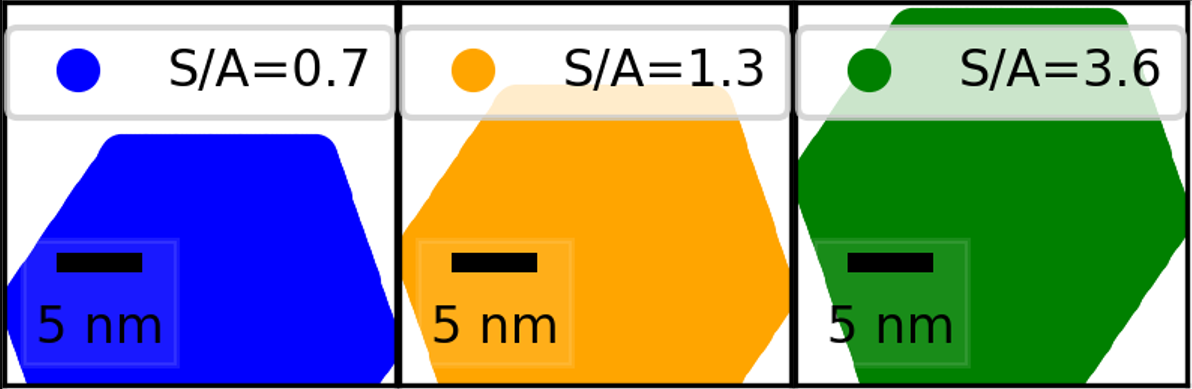}\\
  \includegraphics[width=\linewidth]{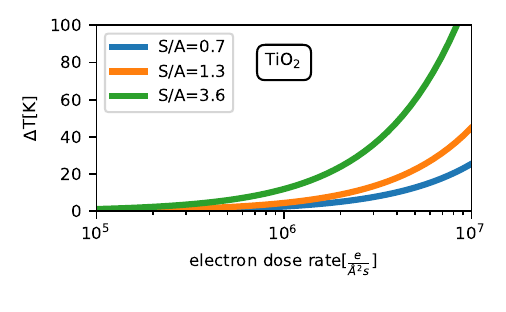}\\
  \includegraphics[width=\linewidth]{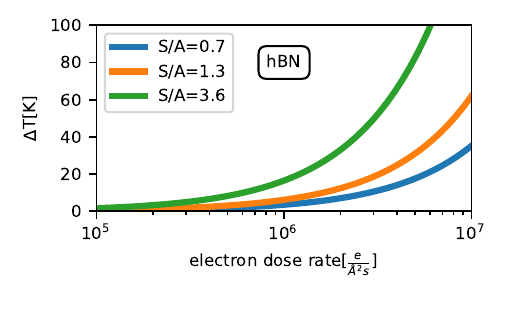}
  \caption{Temperature increase as a function of electron dose rate for small (left) and medium sized (right) gold nanoparticles on \ch{TiO2} substrate at beam energy of $\SI{80}{\kilo\electronvolt}$. Curves are shown for three different ratios of $S/A$ corresponding to the nanoparticles shown in the top.}
    \label{fig:result_small_big}
\end{figure}

Now the temperature increase as a function of dose rate can be calculated, it depends on the area ratio $S/A$, which can vary by more than an order of magnitude depending on the wetting angle, see Figure \ref{fig:SAratio} in the SOI.  Figure \ref{fig:result_small_big} shows the temperature increase as a function of dose rate for \ch{TiO2}-supported nanoparticles of different shape and size.

The heating also depends on the substrate (through $k$) and on the beam energy (through $\lambda$).  This is illustrated in Figure \ref{fig:result_small_big}, showing that reducing the beam energy leads to increased heating, as the mean free path scales as $\lambda \propto E_{\text{beam}}$ \cite{Iakoubovskii2008ThicknessSpectroscopy}.  We also see that the heating is more pronounced on hBN than on \ch{TiO2}, due to the lower $k$ value, which is due to the larger ratio between the masses of the atoms in the nanoparticle and the substrate, leading to a larger mismatch in phonon frequencies.

\begin{figure}[tbp]
    \includegraphics[width=\linewidth]{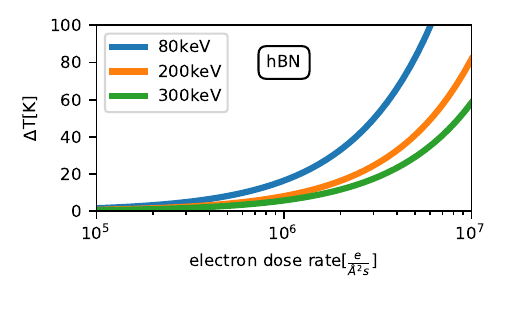}\\
    \includegraphics[width=\linewidth]{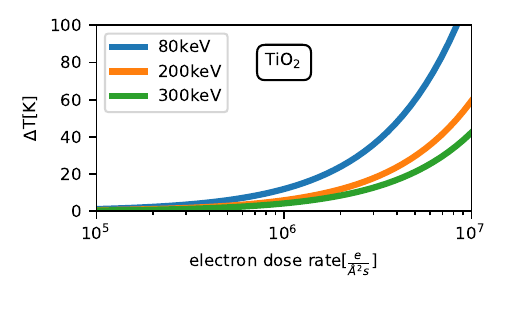}
    \caption{Comparison of the heating of a gold particle with $S/A=3.6$ supported by hBN (left) and \ch{TiO2} (right) for different beam energies, showing both that heating increases with decreasing beam energy, and that the heating is significantly larger on hBN. In the case of beam energies other than $\SI{300}{\kilo\electronvolt}$ or $\SI{80}{\kilo\electronvolt}$, the corresponding $\lambda$ values were taken from \cite{Shinotsuka2015CalculationsAlgorithm}. }
    \label{fig:my_label}
\end{figure}

It should be noted that while increasing the beam energy decreases the heating, it will increase other kinds of beam damage, in particular knock-out damage from Rutherford scattering.  The maximal energy that can be transferred to an atom by Rutherford scattering is \cite{reimer2008springer}: 
\begin{equation}
    E_{max} = 2 E_{\text{beam}} \frac{E_{\text{beam}} + 2 m_e c^2}{M c^2}
\end{equation}
where $m_e$ is the electron rest mass, $M$ the mass of the atom, and $c$ the speed of light.  For gold this gives a maximal energy transfer of \SI{4.3}{\electronvolt} at ${\text{beam}} = \SI{300}{\kilo\electronvolt}$, which is slightly larger than the minimal energy required to remove atoms from gold nanoparticles, reported as \SI{3.8}{\electronvolt} \cite{EGERTON201485}.

\section{Conclusions}

Equivariant Neural Network potentials can be successfully fitted to DFT simulations of small supported nanoparticles, and an ensemble of such potentials can be used to gauge the accuracy of the potentials.  This allows us to extrapolate to much larger systems that cannot be described by DFT, while still having a measure of the error in the predicted energy and forces.  We used these potentials to perform molecular dynamics simulations of the heat transfer coefficients $k$ between the nanoparticle and the supporting substrate.  We found that $k$ depends both on the chemical composition of the substrate and on its crystal structure, with a factor of two between the values found for a gold nanoparticle on rutile and anatase \ch{TiO2}.  We use EELS to measure the mean free path in the nanoparticles, and the average energy lost by scattered electrons.  This finally enables us to predict the heating incurred in nanoparticles of different size and shape as a function of the support material, the beam energy and the beam intensity.

\section{Methods}

\subsection{Fitting the Machine Learning potential}
\label{sec:fittingML}

All DFT calculations were performed using the GPAW DFT calculator \cite{Enkovaara2021GPAW} in the Atomic Simulation Environment \cite{Bahn2002AnCode,Larsen2017TheAtoms}. 
The Projector Augmented Wave (PAW) \cite{Blochl1994ProjectorMethod} method was used to represent the wave functions near the nuclei, and the smoothed wave functions were then represented on a plane wave basis with a cutoff energy of $\SI{500}{\electronvolt}$.  The PBE approximation was used for the exchange-correlation energy \cite{Perdew1996a}.  As all systems were large, the Brillouin zone was sampled with a single $k$-point, the $\Gamma$ point. 

To create the data set, several Molecular Dynamics (MD) simulations were carried out at different temperatures using DFT. Then, we trained a NequIP potential with these configurations. All networks were trained for 200 epochs with a batch size of 7, a learning rate of 0.005, and the Adam optimizer. The loss coefficients for forces and energies were the same. The training set consisted of 2700 DFT calculations, the validation set had 700, and the rest were used as test set.  The machine learning potential generated using this method was used to run MD simulations until configurations were reached that were sufficiently far from the training data to lead to bad predictions.  These bad predictions were determined either by the MD simulation crashing, or by forces appearing that were absurdly large (exceeding $\SI{100}{\electronvolt\per\angstrom}$).  The last configurations before this happened were then recalculated with DFT and added to the training set.  A few iterations of this method led to stable machine learning potentials.  We find that including some configurations with very large forces is essential to properly describe medium-sized forces, see section A.2 in the supplementary information.  The exact same procedure was then followed to train the Au on hBN potential, except that here van der Waals forces were taken into account in the DFT calculations by using the DFT-D3 method \cite{grimme2010consistent}.

Hyperparameters were optimized to obtain the best possible results, within the constraints set by computational resources.  The main hyperparameters to optimize are the cutoff radius, and the number of quantities in each neural network layer for different $l$, where $l$ represent the character of the E(3)-equivariant quantities, such that $l=0$ is scalars and pseudoscalars, $l=1$ is vectors and pseudovectors, and higher $l$ are quantities transforming as spherical harmonics with the respective $l$.  Here, we limit ourselves to $l \le 1$.

\begin{figure}[tbp]
    \centering
    \includegraphics[width=\linewidth]{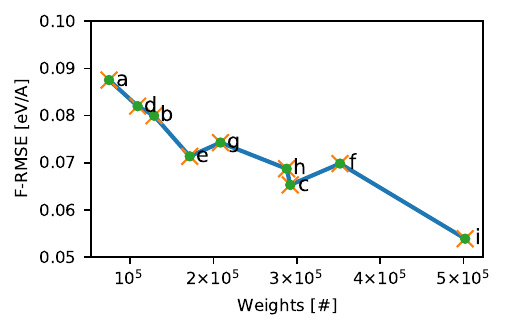}\\
    \includegraphics[width=\linewidth]{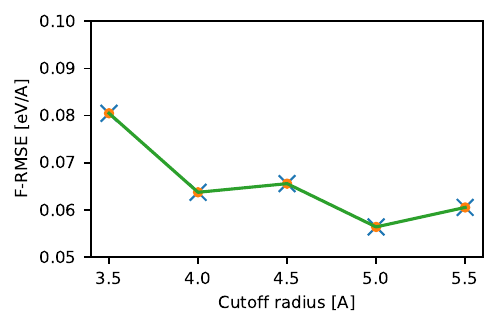}
    \caption{Top panel shows RMSE error of the predicted forces as a function of the number of weights in the neural network.  The letters refer to the combinations of scalars and vectors shown in Table \ref{tab:numberofweights}. Lower panel The same error as a function of cutoff radius.}
    \label{fig:rmse-vs-weights}
\end{figure}

We optimize the number of scalars and vectors in the layers versus the computational complexity of the model.  Table \ref{tab:numberofweights} shows the numbers used and the resulting number of weights in the neural network that need to be optimized, and Figure \ref{fig:rmse-vs-weights}(a) shows the resulting error in the forces predicted on the validation set.  In all cases half the quantities were with even parity, and half with odd (i.e. same number of scalars and pseudoscalars, and same number of vectors and pseudovectors); and in all cases the number of interaction blocks was set to 4.  The best network is the largest that we can comfortably handle, with 64 scalars and 64 vectors, although reducing the number of vectors gives a network with almost the same performance. 
\begin{table}[tbp]
    \centering
    \begin{tabular}{|c|c|c|c|}
        \hline
         \multicolumn{1}{|r|}{Vectors} & 16 & 32 & 64 \\
         \multicolumn{1}{|l|}{Scalars\hspace*{1.5cm}}& & & \\
         \hline
         16 & $74^a$ & $128^b$ & $292^c$  \\
         32 & $109^d$ & $171^e$ & $351^f$  \\
         64 & $208^g$ & $287^h$ & $501^i$  \\
         \hline
    \end{tabular}
    \caption{Number of trainable weights (in thousands) as a function of the number of scalars and the number of vectors.  Half of these have each parity, so half the scalars are actually pseudoscalars and half the vectors are pseudovectors.  The letters are used to label the data points in figure \ref{fig:rmse-vs-weights}(a).}
    \label{tab:numberofweights}
\end{table}

One of the most important hyperparameters in this kind of networks is the cutoff radius, we optimize that in Figure \ref{fig:rmse-vs-weights}(b) where we find that a cutoff radius of \SI{5.0}{\angstrom} is optimal.  It should be noted that the actual range of the potential is larger than the cutoff radius: in each interaction block, the atoms gather information about their neighbors inside the cutoff radius, in the second block the atoms thus have indirect access to information about positions up to two times the cutoff radius, etc.

The machine learning potentials can now be used in molecular dynamics
to model the heat transfer between nanoparticle and support.  For both
types of supports, an potential was fitted as described above.
Initially, separate systems were created for the nanoparticle and the
support.  Both the nanoparticle and the substrate were thermalized to
600K and 300K, respectively, using 15 ps of Langevin dynamics with a
friction parameter of 0.005.  The substrate and nanoparticle are then
brought into thermal contact, and Velocity Verlet dynamics, which
preserves the total energy, is performed for 200 ps.  A time-step of 2
fs is used for both Langevin and Velocity Verlet dynamics.  The Atomic
Simulation Environment (ASE) \cite{Bahn2002AnCode,Larsen2017TheAtoms}
is used for all simulations.

\begin{figure*}[t]
    \includegraphics[width=.49\linewidth]{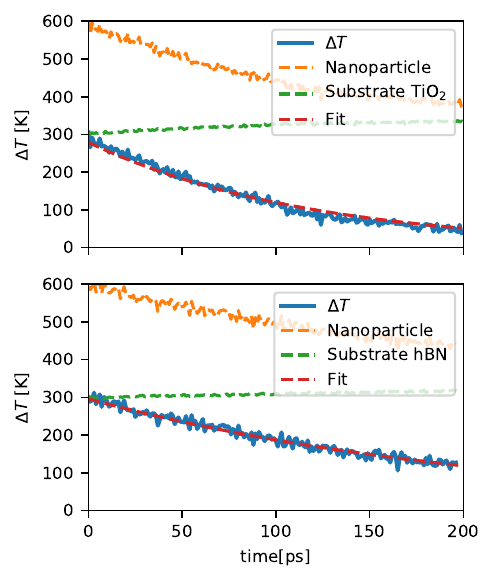}\hfill
    \includegraphics[width=.49\linewidth]{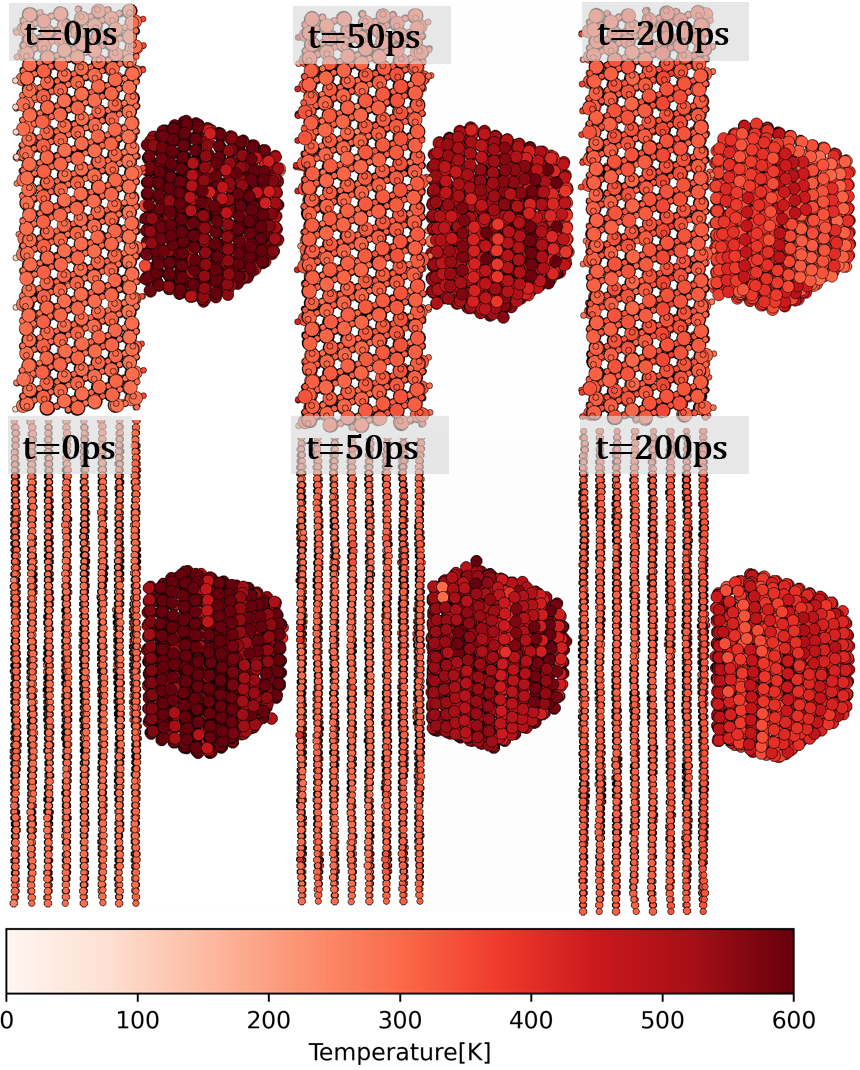}
    \caption{Temperature evolution of nanoparticles and substrate for an Au nanoparticle supported by either \ch{TiO2} (top row) or hBN (bottom row).  The left column shows the temperatures, the right shows the atomic structures with atoms colored according to their temperature, defined as the kinetic energy averages over .  The nanoparticle has $C_p = 2.80\cdot 10^{-4}[\si{\eV\per atom \cdot K}]$ and from the fitted temperature we get $k_{\text{TiO}_2} = 0.54\cdot 10^{-3}[\si{\eV\per\nm\squared\per\ps\per\K}]$ and $k_{\text{hBN}} = 0.29\cdot 10^{-3}[\si{\eV\per\nm\squared\per\ps\per\K}]$.}
    \label{fig:TempFitTio2hBN}
\end{figure*}

Figure \ref{fig:TempFitTio2hBN} shows typical temperature fits  together with snapshots of the systems color-coded by the instantaneous temperature of the atoms, defined from their kinetic energy: $E_{\text{kin}} = \frac32 k T$.  It is seen that the temperature gradient is small within the nanoparticle and within the support, compared to the temperature difference between nanoparticle and support.  From the fits, the time constant $\frac{C c_p}{k A}$ can be extracted, and from there the heat transmission coefficient $k$ is determined.  Clearly, some variability in $k$ is expected (see e.g.\ Figure \ref{fig:TempFitSizes} in the supplementary information), and a table of extracted parameters is available as Table \ref{tab:params}.

\begin{table}[tb]
    \centering
    \footnotesize
    \begin{tabular}{c|c|c|c|c}
         \# & V & A  & $c_p$  & $k$ \\
         \# & [\si{\nm\cubed}] & [\si{\nm\squared}] &  [\si{\eV\per K}] &  [\si{\eV\per\nm\squared\per\ps\per\K}]\\
         \hline
         rut(101)& 3.956 & 2.63  & 2.96$\cdot 10^{-4}$ &1.08$\cdot 10^{-3}$\\
         rut(101) & 7.901 & 3.45  & 2.85$\cdot 10^{-4}$ & 1.12$\cdot 10^{-3}$ \\
         rut(101) & 14.624 & 3.38  & 2.83$\cdot 10^{-4}$ & 1.00$\cdot 10^{-3}$ \\
         rut(101) & 20.338 & 6.52 & 2.80$\cdot 10^{-4}$ & 0.94$\cdot 10^{-3}$ \\

         ana(101) & 3.956 & 2.63 & 2.96$\cdot 10^{-4}$ &0.56$\cdot 10^{-3}$\\
         ana(101) & 7.901 & 3.45 & 2.85$\cdot 10^{-4}$ & 0.54$\cdot 10^{-3}$ \\
         ana(101) & 14.624 & 3.38 & 2.83$\cdot 10^{-4}$ & 0.43$\cdot 10^{-3}$ \\
         ana(101) & 20.338 & 6.52 & 2.80$\cdot 10^{-4}$ & 0.54$\cdot 10^{-3}$ \\

         hBN & 3.956 & 2.63 & 2.96$\cdot 10^{-4}$ &0.44$\cdot 10^{-3}$\\
         hBN & 7.901 & 3.45 & 2.85$\cdot 10^{-4}$ & 0.39$\cdot 10^{-3}$ \\
         hBN & 14.624 & 3.38 & 2.83$\cdot 10^{-4}$ & 0.38$\cdot 10^{-3}$ \\
         hBN & 20.338 & 6.52 & 2.80$\cdot 10^{-4}$ & 0.29$\cdot 10^{-3}$ \\

    \end{tabular}
    \caption{Extracted heat transfer coefficients $k$ and heat capacities per atom ($c_p$) for Au supported on \ch{TiO2} and hBN.}
    \label{tab:params}
\end{table}

\subsection{Measuring transfer of energy}
\label{sec:experimental}

The sample was made by ultrasonicating hexagonal boron nitride (hBN) in ethanol for 5 minutes. The suspension was then drop-cast on a DENSsolutions Wildfire through-hole heater chip. The chip was left to dry and subsequently coated with 3 nm gold in a Quorom Q150T sputter coater. The sample was inserted in the microscope and heated to 500°C for 30 minutes resulting in a break-up of the Au film into particles (see Figure \ref{fig:experimental}).

Drift-corrected electron energy-loss line spectra were acquired using a FEI Titan ETEM in scanning transmission electron microscopy mode (STEM). Data was acquired at both 300 kV and 80 kV. The entrance aperture of the spectrometer was 2.5 mm giving a collection semi-angle of 7.4 mrad and the acquisition time was 10 ms in each spot.

\begin{figure*}[tbp]
    \begin{subfigure}{0.49\linewidth}
        \centering
        \stackinset{l}{5pt}{t}{5pt}{(a)}{\includegraphics[width=\linewidth]{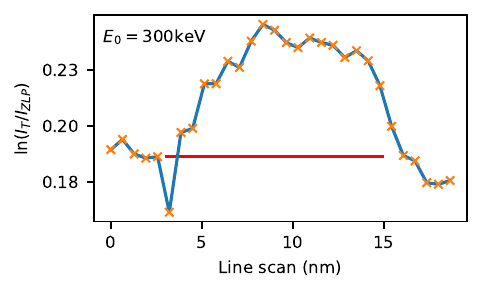}}
    \end{subfigure}
    \hfill
    \begin{subfigure}{0.49\linewidth}
        \centering
        \stackinset{l}{5pt}{t}{5pt}{(b)}{\includegraphics[width=\linewidth]{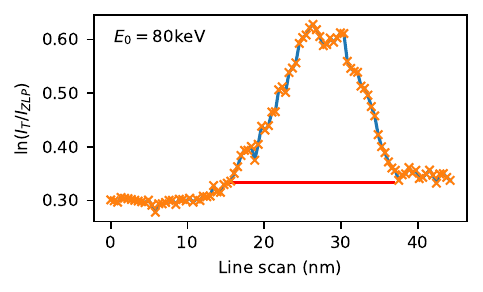}}
    \end{subfigure}

    \begin{subfigure}{0.49\linewidth}
        \centering
        \stackinset{l}{5pt}{t}{5pt}{(c)}{\includegraphics[width=\linewidth]{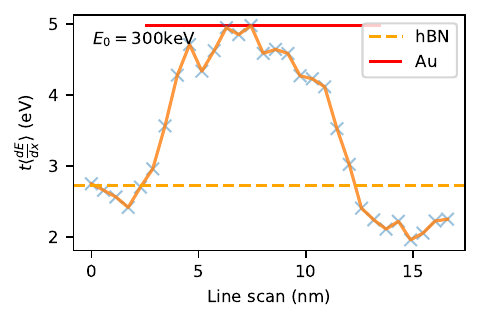}}
    \end{subfigure}
    \hfill
    \begin{subfigure}{0.49\linewidth}
        \centering
        \stackinset{l}{5pt}{t}{5pt}{(d)}{\includegraphics[width=\linewidth]{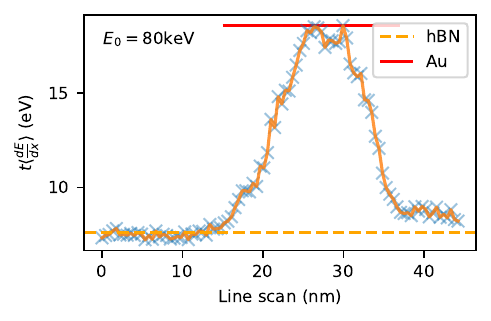}}
    \end{subfigure}

    \begin{subfigure}{0.49\linewidth}
        \centering
        \stackinset{l}{5pt}{t}{5pt}{(e)}{\includegraphics[width=\linewidth]{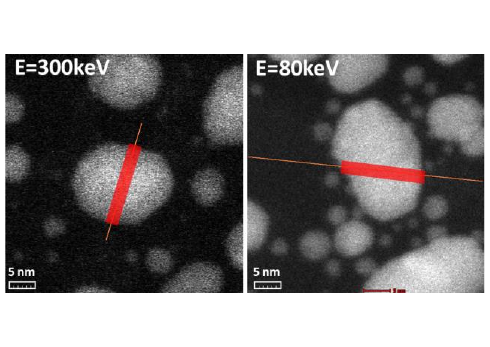}}
    \end{subfigure}
    \hfill
    \begin{subfigure}{0.49\linewidth}
        \centering
        \stackinset{l}{5pt}{t}{5pt}{(f)}{\includegraphics[width=\linewidth]{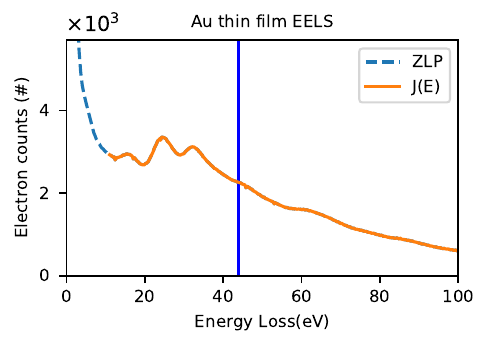}}
    \end{subfigure}
    \caption{The mean free path $\lambda$ is calculated from the relative thickness obtained using the log ratio method on the left column using EELS Line scans ($T = \frac{t}{\lambda}=\ln\frac{I_T}{I_{ZLP}}$).  Panel (a) and (b) show data for Au-NP supported by hBN at different beam energies (300 keV and 80 keV). The red line highlights the nanoparticle in the line scan. The values for $\lambda$ are  $\SI{104}{\nano\meter}$ and $\SI{37}{\nano\meter}$ for $\SI{300}{\kilo\electronvolt}$ and $\SI{80}{\kilo\electronvolt}$ respectively.  Panel (c) and (d) show the average energy lost by the electrons in the beam, from the estimated thickness of the nanoparticle the energy loss per distance is calculated, we find that values for $\langle \frac{dE}{dx}\rangle_{NP}$ are $\approx \SI{0.4}{\eV\per\nm}$ and $\approx \SI{1.0}{\eV\per\nm}$ for the two beam energies.  Panel (e) shows the line scans across the particles used to take the data.  Panel (f) visualizes the calculation of average energy loss per inelastic collision within a thin Au film. This calculation is based on the determination of the expected energy value within the low-loss spectrum ($J(E)$), as indicated by the orange line. The solid blue line represents the estimated expected value ($\langle E \rangle \approx 44 eV$) and the blue dotted line the Zero Loss Peak (ZLP) region. Data from EELS.info \cite{eelsinfo}.}
    \label{fig:findlambda}
\end{figure*}

The mean free path of the electrons in the sample ($\lambda$) is measured by calculating the relative thickness $T = \frac{t}{\lambda}$ from EELS spectra of Au-nanoparticles on hBN, assuming approximately hemispherical shape of the nanoparticle, see Figure \ref{fig:findlambda}.  We find $\lambda$ as
\begin{equation}
    \lambda = \frac{R_{\text{est}}}{H_{\text{center}} - H_{\text{background}}}
\end{equation}
where $R_{\text{est}}$ is the thickness of the nanoparticle estimated from its observed radius, and $H_{\text{center}}$ and $H_{\text{background}}$ are the calculated $t/\lambda$ ratios for the center of the nanoparticle and the background, respectively, calculated as $H = t / \lambda = \ln \left(I_T / I_\text{{ZLP}}\right)$, where $I_T$ is the total intensity of the spectrum, and $I_{\text{ZLP}}$ is the intensity of the zero-loss peak, i.e. the intensity that is transmitted without being scattered.

In our EELS data processing methodology, all EELS spectra undergo alignment to the zero-loss peak, which is positioned at energy zero. We subsequently define the zero-loss peak region as spanning the energy interval ($-E_{th}$, $E_{th}$), where $E_{th}$ represents the calculated elastic scattering threshold energy of the EELS spectrum. Furthermore, the low-loss region is defined as spanning the range $(E_{th}, 110eV)$. The alignment and threshold energy calculations are carried out using the EELS module integrated into the Hyperspy package \cite{de2022hyperspy}.

While there is some uncertainty in these measurements, it gives
$\lambda$ at least within a factor two.    The
average mean free path value is determined from the data in figure \ref{fig:findlambda} to be
$\lambda = \SI{104}{\nano\meter}$ for a beam energy of
\SI{300}{\kilo\electronvolt} and $\lambda = \SI{37}{\nano\meter}$ at \SI{80}{\kilo\electronvolt} .  This is in good agreement with the
theoretically determined value of $\lambda = \SI{74}{\nano\meter}$ for
a $\SI{200}{\kilo\electronvolt}$ beam
\cite{Shinotsuka2015CalculationsAlgorithm} on bulk gold, as the mean free
path is expected to increase with increasing energy.  Similar values are found from two different experiments, see Figure
\ref{fig:findlambda-suppl} in the supplementary information.

Additionally, we determined the average inelastic energy loss, denoted as $\langle E \rangle$, within a thin gold (Au) film through electron energy-loss spectroscopy (EELS) measurements. To obtain this value, we calculated the expectation value of the spectrum within the low loss region, yielding a value of approximately $44$ eV.

An alternative way to estimate the power energy absorbed by the nanoparticle due to the electron beam irradiation previously used in \cite{zheng2009nanocrystal}, is:
\begin{equation}
    \dot Q_{in} = S D t \langle \frac{dE}{dx} \rangle
\end{equation}
Here, $t \langle \frac{dE}{dx} \rangle$ represents the mean energy loss of electrons per unit distance traveled through a material of thickness $t$. By calculating the energy's expectation value from the EELS signal, including the Zero Loss Peak (ZLP), along the line scan, we obtain the value of $t \langle \frac{dE}{dx} \rangle$. The contribution from the substrate is quantified as $t_{hBN} \langle \frac{dE}{dx} \rangle_{hBN}$.
 
The relationship between $t \langle \frac{dE}{dx} \rangle$ and $t_{hBN} \langle \frac{dE}{dx} \rangle_{hBN}$ is given by the following equation:

\begin{align}
t \langle \frac{dE}{dx} \rangle &= t_{NP} \langle \frac{dE}{dx} \rangle_{NP} + t_{hBN} \langle \frac{dE}{dx} \rangle_{hBN} 
\end{align}

By knowing the thickness of the nanoparticle (approximated as a hemisphere) and taking the values of $t_{hBN} \langle \frac{dE}{dx} \rangle_{hBN}$ just before and after the nanoparticle, we can determine $\langle \frac{dE}{dx} \rangle_{NP}$. Figure \ref{fig:findlambda}(c-d) shows $t\langle \frac{dE}{dx} \rangle$ at an electron energy of 300 keV and 80 keV, which is approximately $\langle \frac{dE}{dx} \rangle_{NP}\approx \SI{0.4}{\eV\per\nm}$ and  $\approx \SI{1.0}{\eV\per\nm}$ respectively. This value is compared to the theoretical value of around $\approx \SI{0.26}{\eV\per\nm}$ for bulk Au at 300 keV, calculated using the Bethe equation \cite{zheng2009nanocrystal}. 

Upon comparing the two previously described methods for estimating heat input into the nanoparticle, we observed a substantial level of agreement, with a percentage agreement of approximately $95\%$ for the 300 keV electron beam energy and approximately $85\%$ for the 80 keV energy. As a result, we have chosen to adopt the method described by equation \ref{eq:qout}. We have utilized these values in a qualitative manner, taking into consideration the variability of the electron mean free path, average energy, and relative thickness under different experimental conditions  \cite{zhang2012local}. However, it is important to note that our specimen has a significantly smaller thickness compared to the electron mean free path ($\lambda\gg t$).

\section{Acknowledgments}

The authors acknowledge financial support from the Independent
Research Fund Denmark (DFF-FTP) through grant no.~9041-00161B.


\bibliography{cleanedreferences}

\clearpage
\appendix

\onecolumn

\section{Supplementary Online Information}
\setcounter{figure}{0}
\renewcommand{\thefigure}{SOI-\arabic{figure}}
\input{supplementary}

\end{document}

%% file: supplementary.tex
\subsection{Learning curve}

\begin{figure}
    \centering
    \includegraphics{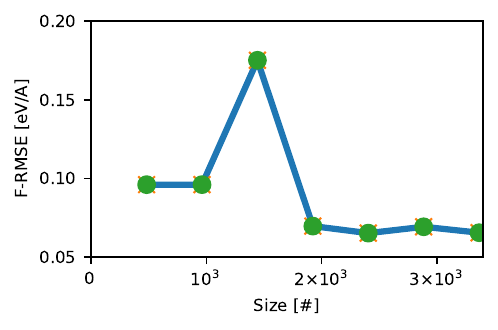}
    \caption{Learing curve: The RMSE force error in the validation set as a function of the training set size (the validation size was the same for all the models).}
    \label{fig:learningcurve}
\end{figure}

Figure \ref{fig:learningcurve} shows how the RMSE error of the forces depend on the training set size.  The curve is obtained by randomly picking subsets of the full training set, and shows that beyond 2000 configurations in the training set no further improvement is seen.

\subsection{Surface area ratio}

\begin{figure}[h]
    \centering
    \includegraphics{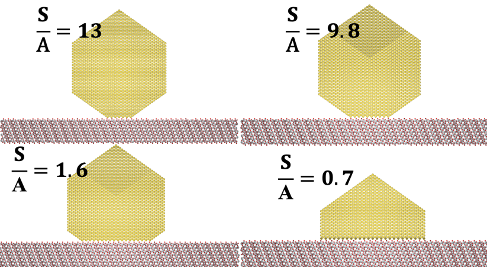}
    \caption{Depending on the wetting angles, the $S/A$ ratios can vary by an order of magnitude or more.}
    \label{fig:SAratio}
\end{figure}

The ratio between the contact area for thermal conductivity ($A$) and the  transverse area presented to the beam may vary by more than an order of magnitude, see Fig.~\ref{fig:SAratio}.

\subsection{Importance of unrealistically large forces when training}
\label{sec:largeforces}

\begin{figure*}
    \includegraphics[width=0.6\linewidth]{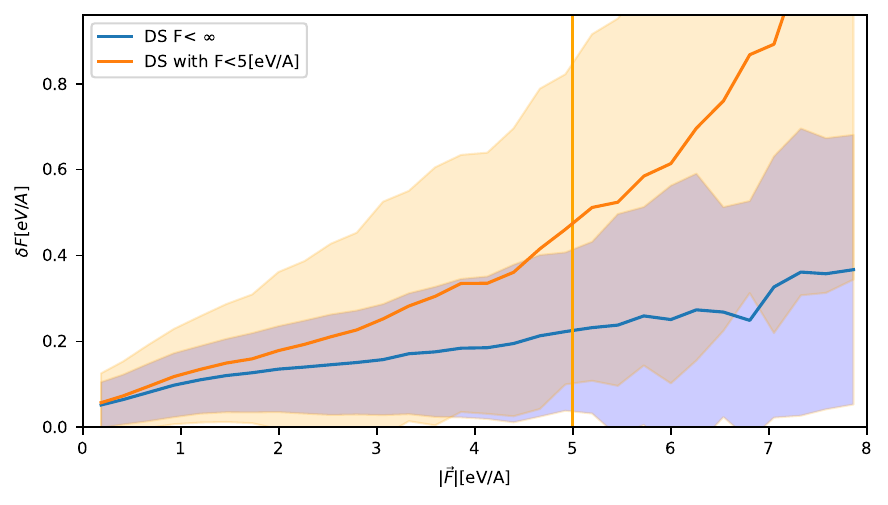}\\
    \includegraphics[width=0.49\linewidth]{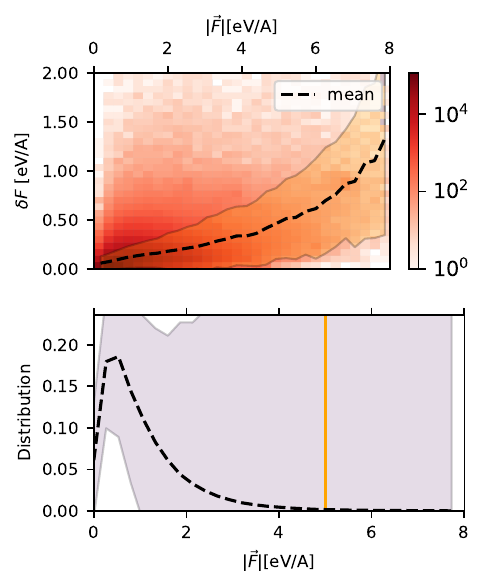}\hfill
    \includegraphics[width=0.49\linewidth]{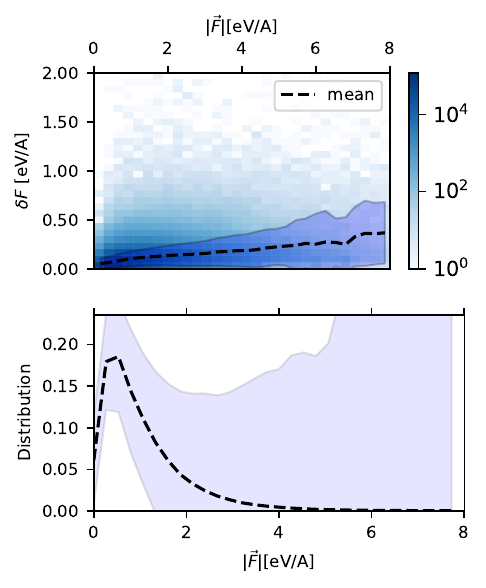}
    \caption{The top panel of the graph illustrates the performance of two potentials: the orange curve represents the potential trained with a training set containing forces below 5eV/A, while the blue curve corresponds to a potential trained without restrictions on the magnitude of forces. Both potentials were trained with training set of identical size. Error bars represent the the standard deviation of $\delta F$ within each column bin. A vertical line is placed at $|\vec{F}|=5$eV/A to indicate the forces cutoff threshold.
    The lower panel displays the histogram of $\delta F$ as a function of force magnitude. Additionally, we provide a visualization of the forces distribution within our testing dataset. To aid interpretation, error bars are included to visually emphasize the standard deviation of $\delta F$ within each column bin. This illustration offers valuable insights into how error variability correlates with different force magnitudes}
    \label{fig:largeforces}
\end{figure*}

In principle, the potential should be able to extrapolate the configuration with relatively large forces on the atoms without any problem. In the figure \ref{fig:largeforces} we can see how diverse is our training set in the magnitude of the force, most of the dataset lies with relative small forces (between 0--2 eV/A), however the error versus magnitude of the force is  a relative flat curve (see lower panel figure \ref{fig:largeforces}) which means that the model is also able to predict large forces; the errors in large forces (between 3--5eV/A) are around twice the average of the error in the small forces(between 0--1eV/A). 

Given that we are interested in modelling systems that on average have relatively small forces  (between 0--5 eV/A), the majority of our training set lie in this range. However, we trained a model where the training set only contains forces below 5 eV/A; we can see that the results are significantly worse than for a model trained with the same training set size but not excluding the small fraction of configurations with large forces (Figure \ref{fig:largeforces}, lower panels).  We speculate that this is because it is important for the neural network to see atomic configurations that would not be well described by a harmonic approximation near a local energy minimum.

\clearpage
\subsection{Determination of the heat transfer coefficient}

\begin{figure*}[hb]
    \adjustbox{valign=T}{\includegraphics[width=0.5\linewidth]{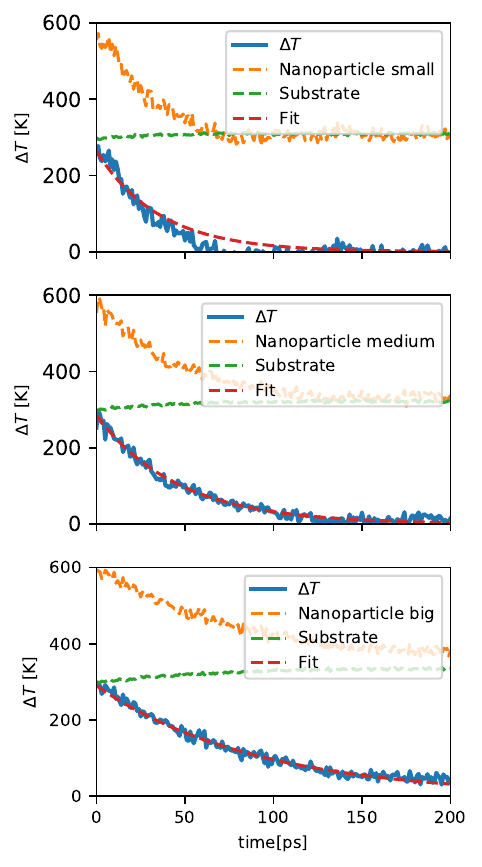}}
    \adjustbox{valign=T}{\includegraphics[width=0.48\linewidth]{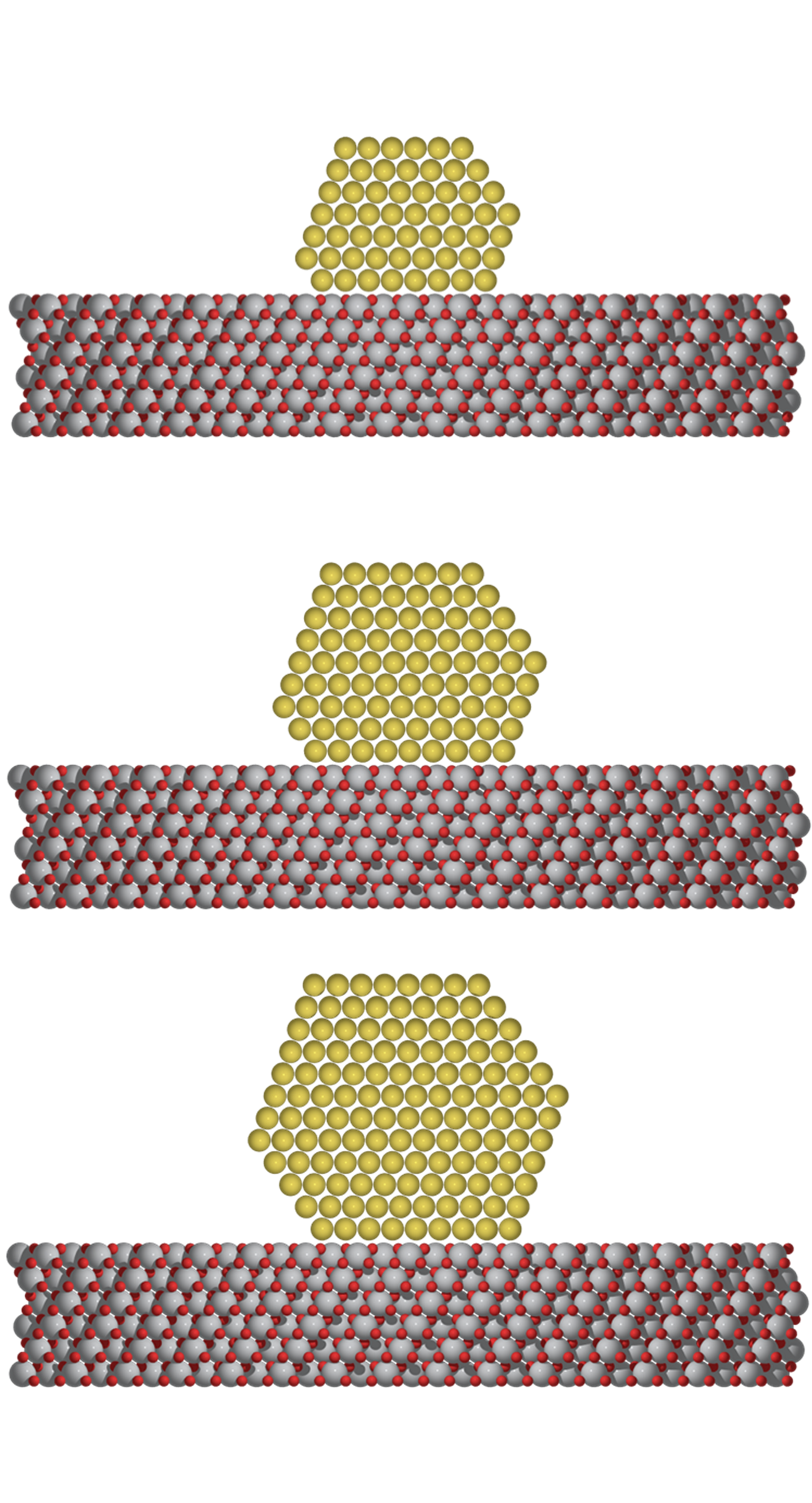}}
    \caption{Temperature fits of three different Au nanoparticles with different volume and contact area. The values for $k$ are $k = \SI{1.08e-3}{\electronvolt\per\nm\squared\per\pico\second}$, $k = \SI{1.12e-3}{\electronvolt\per\nm\squared\per\pico\second}$, and $k = \SI{1.0e-3}{\electronvolt\per\nm\squared\per\pico\second}$ respectively.}
    \label{fig:TempFitSizes}
\end{figure*}

Figure \ref{fig:TempFitSizes} shows the determination of the heat transfer coefficient $k$ for three differently sized gold nanoparticles on rutile \ch{TiO2}, corresponding to the three first rows in Table \ref{tab:params}.

\clearpage

\subsection{Overview image of a sample}

\begin{figure}[htbp]
    \centering
    \includegraphics[width=8cm]{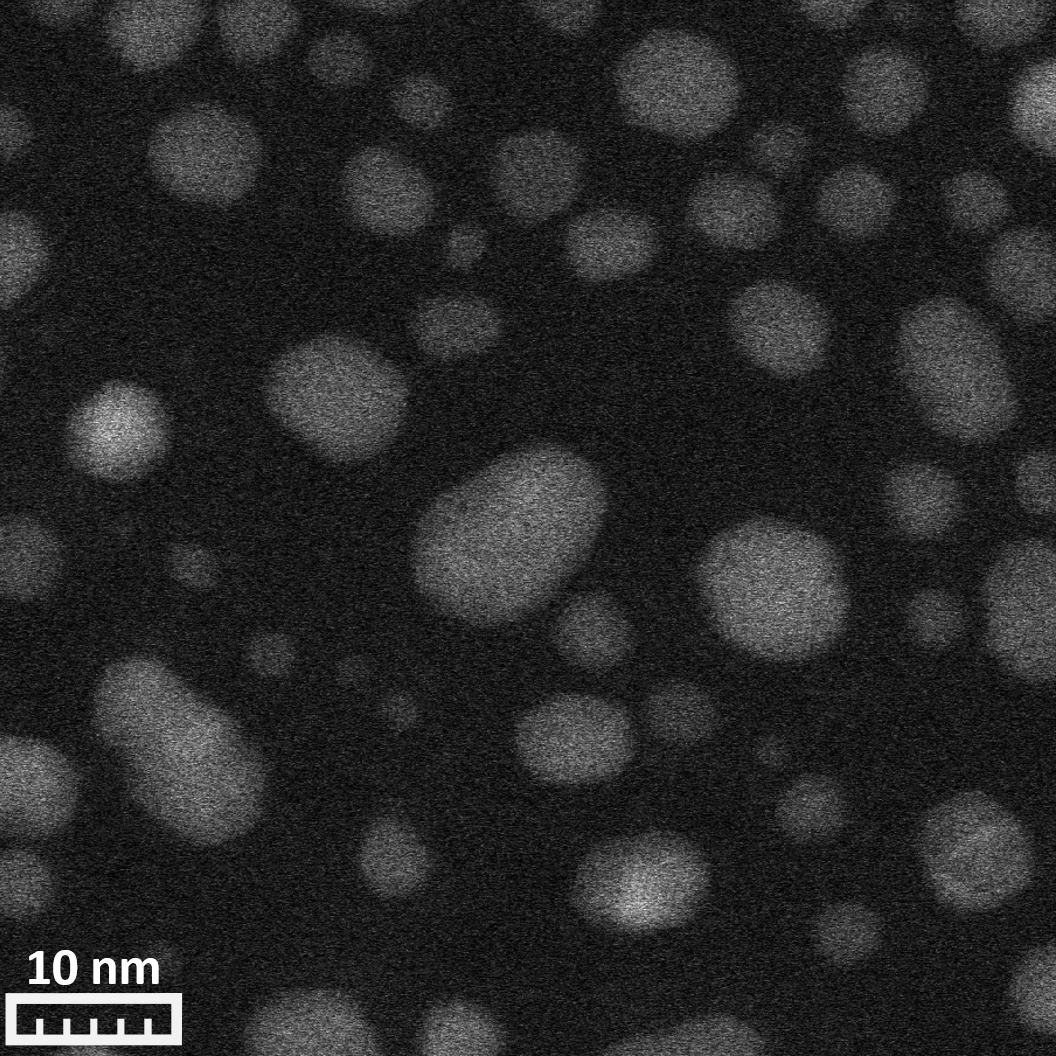}
    \caption{STEM image of gold nanoparticles supported by hBN at 300keV}
    \label{fig:experimental}
\end{figure}

Figure \ref{fig:experimental} shows an overview image of one of samples produced displaying gold nanoparticles on a hBN substrate, as described in the main text.

\clearpage
\subsection{Additional determinations of the electron mean free path}

\begin{figure*}[hb]
    \includegraphics[width=0.49\linewidth]{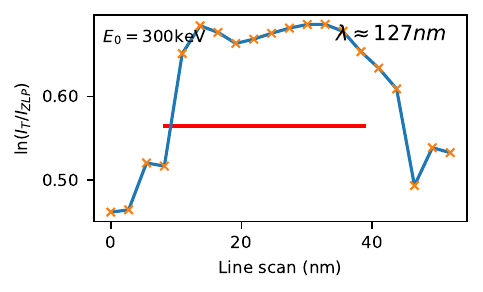}\hfill
    \includegraphics[width=0.49\linewidth]{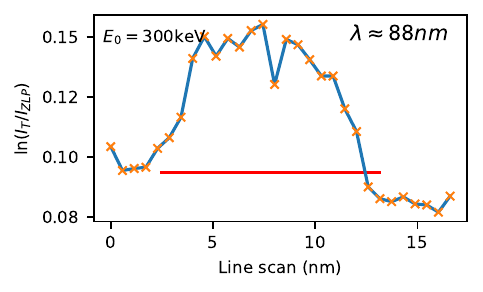}\\
    \includegraphics[width=0.49\linewidth]{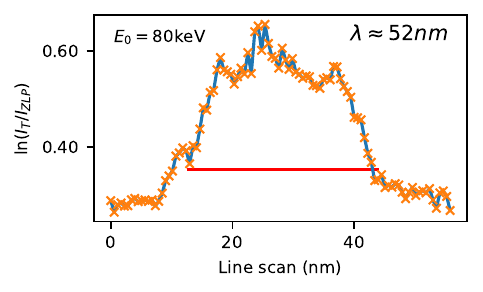}\hfill
    \includegraphics[width=0.49\linewidth]{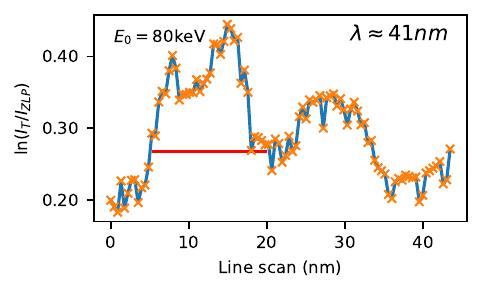}

    \caption{Additional determination of the electron mean free path $\lambda$.  See figure \ref{fig:findlambda} in the main text.}
    \label{fig:findlambda-suppl}
  \end{figure*}
  
The electron mean free path was determined from the most regularly shaped nanoparticles in Figure \ref{fig:findlambda}.  Similar results are found here from two additional nanoparticles at each beam energy, see Figure \ref{fig:findlambda-suppl}.

\clearpage

\subsection{Electron Energy Loss Spectrum}

To determine the average energy loss per inelastic scattering event in Electron Energy Loss Spectroscopy (EELS), we analyze the inelastic low-loss region of the EELS spectrum, typically ranging from 5 to 110 eV and the Zero Loss Peak (ZLP) ranging from $-E_{th}$ to $-E_{th}$. The signal is treated as follows:

$$
\text{Signal}(E) =
\begin{cases}
ZLP(E), & \text{if } -E_{th}<E < E_{th} \\
J(E), & \text{if } E_{th}<E < 110eV
\end{cases}
$$

\begin{figure}[b]
    \centering
    \includegraphics[width=0.9\linewidth]{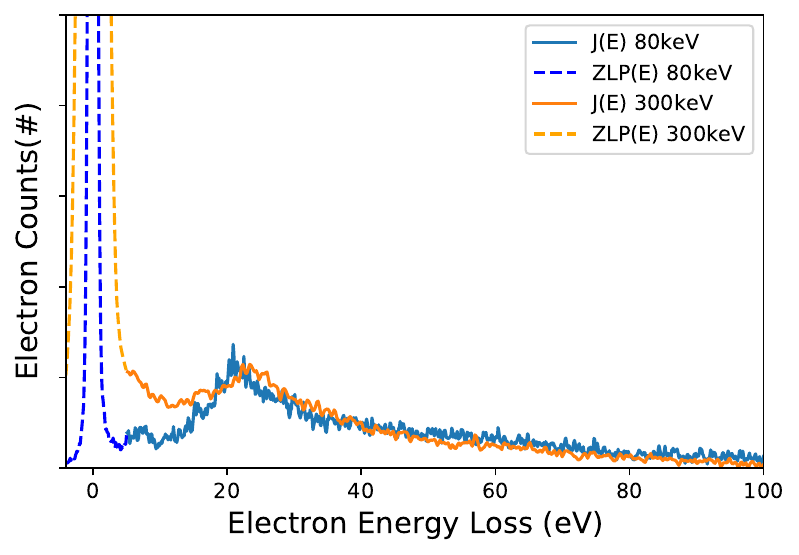}
    \caption{Treatment of the signal, dotted lines are the ZLP region and solid lines are inelastic low-loss region of the EELS }
    \label{fig:zlpje}
\end{figure}

Here, $E_{th}$ represents the elastic scattering threshold energy of the signal. This threshold is determined using the hyperspy package, which provides tools for EELS analysis. See Figure \ref{fig:zlpje}.

